\documentclass{article}

\usepackage{arxiv}

\usepackage[utf8]{inputenc} 
\usepackage[T1]{fontenc}    
\usepackage{hyperref}       
\usepackage{url}            
\usepackage{booktabs}       
\usepackage{amsfonts}       
\usepackage{nicefrac}       
\usepackage{microtype}      
\usepackage{lipsum}
\usepackage{graphicx}
\graphicspath{ {./figures/} }
\usepackage{caption}
\usepackage{subcaption}
\usepackage{amsmath}
\usepackage{todonotes}

\title{Uncertainty Quantification in Seismic Inversion Through Integrated Importance Sampling and Ensemble Methods}

\author{
Luping Qu \\
Earth Resources Laboratory \\
Massachusetts Institute of Technology \\
Cambridge, MA 02139 \\
\texttt{lupingqu at mit.edu} \\
\And
Mauricio Araya-Polo \\
TotalEnergies EP Research \& Technology US LLC \\
Houston, TX, 77002 \\
\texttt{mauricio.araya at totalenergies.com} \\
\And
Laurent Demanet \\
Department of Mathematics and \\
Department of Earth, Atmospheric, and Planetary Sciences \\
Massachusetts Institute of Technology \\
Cambridge, MA 02139 \\
\texttt{ldemanet at mit.edu} 
}

\begin{document}
\maketitle
\begin{abstract}
Seismic inversion is essential for geophysical exploration and geological assessment, but it is inherently subject to significant uncertainty. This uncertainty stems primarily from the limited information provided by observed seismic data, which is largely a result of constraints in data collection geometry. As a result, multiple plausible velocity models can often explain the same set of seismic observations. 
In deep learning-based seismic inversion, uncertainty arises from various sources, including data noise, neural network design and training, and inherent data limitations. 
This study introduces a novel approach to uncertainty quantification in seismic inversion by integrating ensemble methods with importance sampling. By leveraging ensemble approach in combination with importance sampling, we enhance the accuracy of uncertainty analysis while maintaining computational efficiency. The method involves initializing each model in the ensemble with different weights, introducing diversity in predictions and thereby improving the robustness and reliability of the inversion outcomes. Additionally, the use of importance sampling weights the contribution of each ensemble sample, allowing us to use a limited number of ensemble samples to obtain more accurate estimates of the posterior distribution. Our approach enables more precise quantification of uncertainty in velocity models derived from seismic data. By utilizing a limited number of ensemble samples, this method achieves an accurate and reliable assessment of uncertainty, ultimately providing greater confidence in seismic inversion results.
\end{abstract}

\section{Introduction}

Uncertainty Quantification (UQ) is the process of identifying, quantifying, and managing the uncertainties inherent in model predictions and data interpretations. In geophysical problems, UQ plays a vital role in ensuring that decisions based on model outputs are robust and reliable. Specifically, in seismic inversion, UQ is crucial due to the non-uniqueness of solutions —- multiple subsurface models can produce the same seismic data because of the indirect nature of the measurements. The uncertainties have an aleatoric source (due to the intrinsic randomness of the observations), but, in contexts where machine learning is used, also an epistemic source (stemming from incomplete knowledge, or nonunique specification, of the model that predicts the data).  By applying UQ, geophysicists can quantify the range of possible models, assess their likelihood, and make better-informed decisions regarding drilling, exploration, and resource management. This not only reduces the financial risks associated with incorrect predictions but also enhances the overall effectiveness of exploration and production activities.


Across various geophysical problems, traditional and modern Uncertainty Quantification (UQ) methods have evolved to address the specific needs and challenges of each domain. Traditional methods often rely on statistical techniques such as Monte Carlo simulations, which generate multiple realizations of a model to estimate the uncertainty in predictions. For example, in seismic inversion, traditional UQ might involve running numerous inversion scenarios with varied input parameters to explore the possible range of velocity models. While this approach is effective, it can be computationally expensive and may struggle with the high-dimensionality and complexity of modern geophysical problems. To overcome these limitations, a variety of UQ methods have been developed, including Variational Inference (VI)(\cite{hoffman2013stochastic},\cite{zhang2021introduction}), Hamiltonian Monte Carlo (HMC)(\cite{fichtner2019hamiltonian},\cite{liang2023uncertainty}, and \cite{peng20242}), and ensemble techniques(\cite{anderson2009ensemble},\cite{iglesias2013ensemble}). HMC provides a robust theoretical foundation for sampling from complex, high-dimensional distributions but is often hampered by its high computational demands and the need for careful parameter tuning. VI, on the other hand, offers greater computational efficiency compared to sampling-based methods like HMC, though it is limited by approximation biases and the complexities involved in model design. Ensemble methods, such as the Ensemble Kalman Filter (EnKF), are also widely used, with their effectiveness largely depending on the size of the ensemble; a smaller ensemble may fail to adequately represent the true posterior distribution. Each of these methods brings its unique advantages and challenges, contributing to the continuous evolution of UQ techniques in geophysics.

With the advent of Deep Learning, new UQ methods have emerged, offering more sophisticated and efficient ways to quantify uncertainty. In Deep Learning-based seismic inversion, for example, UQ can be achieved through techniques like Bayesian Neural Networks (BNNs) and Monte Carlo Dropout (MC dropout).  The former is rooted in a solid theoretical foundation that can lead to reliable results, but it is computationally costly. The latter can be seen as a cheap approximation to what BNN can do produce. Unfortunately, MC dropout approximations regularly fail to capture the full extent of uncertainty, particularly in complex models where the dropout mechanism might oversimplify the representation of uncertainty. These limitations highlight the need for more efficient and reliable UQ methods in the context of waveform inversion.

Deep Ensemble methods(\cite{fort2019deep},\cite{ganaie2022ensemble}), appear as a relevant alternative to the above described methods. Deep Ensemble methods have the advantage of introducing diversity in predictions, which can lead to more robust uncertainty estimates. However, a small ensemble size may not accurately represent the true uncertainty, leading to biased or incomplete posterior estimates. Conversely, increasing the ensemble size improves accuracy but at the cost of significantly higher computational resources. Therefore, the main challenge lies in finding a way to use a limited ensemble size to accurately estimate the UQ posterior, balancing the trade-off between computational efficiency and the reliability of uncertainty estimates.

In our study, we address this challenge by combining the Deep Ensemble approach with Importance Sampling (\cite{tokdar2010importance}). This combination allows us to not only generate diverse neural network models with different initial weights but also to calculate the weight of each ensemble member according to its contribution to the posterior distribution. By assigning weights to the ensemble samples, we can better understand how each model contributes to the overall uncertainty estimation, leading to a more accurate and computationally efficient UQ process. This approach enhances the reliability of the posterior distribution estimation, even with a limited ensemble size, providing more precise insights into the uncertainties inherent in seismic inversion.

\section{Related Work}

Methods to address UQ for ML review in this work:

\begin{itemize}
    \item Bayesian Neural Networks (BNN) (\cite{hernandez2015probabilistic},\cite{springenberg2016bayesian}, \cite{sengupta2020ensembling}, and \cite{zhang2021bayesian}). BNNs incorporate uncertainty directly into the
model by treating weights as distributions rather than fixed values, providing a probabilistic framework that naturally
captures epistemic uncertainty. Unfortunately, for waveform inversion problems, the application of Bayesian Neural Networks (BNNs) becomes impractical due to the enormous computational demands: Nonparametric BNNs would involve orders of magnitude more parameters than there are weights in the network, while gaussian approximations of weight distributions are far from capturing their highly multimodal nature.

    \item Monte Carlo Dropout (MC Dropout)(\cite{gal2016dropout}, \cite{chen2024combining}), approximates Bayesian inference by randomly
dropping out nodes during training and inference, generating an ensemble of predictions that reflect the model’s
uncertainty. While MC Dropout offers a
computationally cheaper alternative, it comes with its own limitations. MC Dropout mimics Bayesian inference by randomly dropping out nodes during training and inference, but this approach offers no more than an indication of randomness -- it does not compute actual deviations.


    \item Invertible Neural Networks (INN) (\cite{zhang2021bayesian, Sun24}). INNs are bijective function approximators which have a forward mapping and an inverse mapping that is learned by using the same network architecture. Unfortunately, INNs
maximum mean discrepancy values, which measures the mismatch of a pair of distributions represented by sets of samples, is erratic when exposed to few samples. Therefore, even if this approach is one of the most reliable approximators of MCMC-level sampling for epistemic UQ it turns to be computationally costly when a significant number of samples is used.
    
\end{itemize}

\section{Methodology}
\subsection{Uncertainty Quantification in Neural Networks for Inverse Problems}

In the context of operator learning, where the goal is to map inputs to outputs via a neural network, we define the function \( y = g_w(x) \), where \( g_w \) represents a neural network with weights \( w \). In seismic inversion, the inputs \( x \) typically correspond to seismic observation data, while the outputs \( y \) represent the corresponding velocity models.

In the Bayesian setting, the posterior distribution on the weights \( w \) given the training data \( (X, Y) \) is given by
\[
p_{\text{pos}}(w \mid X, Y) \propto p(Y \mid X, w) p_{\text{pri}}(w),
\]
where \( p(Y \mid X, w) \) is the likelihood of observing the outputs \( Y \) given the inputs \( X \) and weights \( w \), and \( p_{\text{pri}}(w) \) is the prior distribution on the weights. For simplicity, the prior \( p_{\text{pri}}(w) \) can be taken as uniform over a finite domain, such as a hypercube.

The likelihood function, assuming Gaussian noise with standard deviation \( \sigma \), takes the form
\[
p(y \mid x, w) \propto \exp\left[ -\frac{\| y - g_w(x) \|^2}{2 \sigma^2} \right],
\]
where \( \| \cdot \| \) denotes the Euclidean norm. The choice of prior function is a crucial component of UQ, although in this work, we consider a uniform prior for simplicity.

The training process minimizes the negative log-posterior, which serves as the loss function:
\[
L(w) = -\log p_{\text{pos}}(w \mid X, Y).
\]

At the testing phase, the predictive distribution for a test input \( x \) (seismic data) is given by
\[
p(y \mid x) = \int p(y \mid x, w) p_{\text{pos}}(w \mid X, Y) \, dw,
\]
and the expected value of a function \( f(y) \) under this predictive distribution is the quantity-of-interest (QoI)
\[
\mathbb{E}[f(y) \mid x] = \int \mathbb{E}[f(y) \mid x, w] p_{\text{pos}}(w \mid X, Y) \, dw.
\]
For instance, by taking \( f(y) = y_i \), one obtains the expected value of the \( i \)-th component of \( y \) (e.g., a specific feature of the velocity model), and by choosing \( f(y) = (y_i - \mathbb{E}[y_i])^2 \), one obtains the variance of that component.

If a set of samples $w_i$ were drawn from the posterior distribution $p_{\text{pos}}(w \mid X, Y)$, one could compute good empirical estimators of the QoI as
\begin{equation}\label{eq:empirical_f}
\mathbb{E}[f(y) \mid x] \approx \frac{1}{N} \sum_{i=1}^N f(g_{w_i}(x)).
\end{equation}
However, generating samples from the weight posterior is computationally intractable in practice.

\subsection{Deep Ensembles for UQ}

 A simple approach for generating the weights $w_i$ is to perform (stochastic) gradient descent on sets of initial weights drawn independently at random, in order to obtain low values of the training loss independently for each set of weights. For simplicity, the formula (\ref{eq:empirical_f}) is still used to estimate the QoI. This strategy is called Deep Ensemble. 

There is no guarantee that the deviations obtained from deep ensembles bear any resemblance to the Bayesian truth, since the distribution that the weights $w_i$ sample is not the weight posterior. In fact, for simple posterior distributions such as log-concave, it is easy to see that the points $w_i$ would converge to the unique maximum a posteriori point, hence would be far from spanning a representative range of weights where the posterior is large.

It stands to reason, however, that for a log-posterior with highly complex landscape, the deep ensemble weights will settle in different basins of attraction, or valleys, in a fairly spread-out manner while at the same time attaining high values of the posterior. In very high dimensions, we can even surmise that the spread of the weights across the different basins dwarfs the width of each of these basins, obviating the need to precisely explore the landscape locally. For second moments (covariances) this behavior would be analogous to the computation of the moment of inertia of a union of faraway objects -- the particular shapes of those objects doesn't matter nearly as much as their overall configuration. For this reason, we still benchmark the Deep Ensemble as a viable candidate for UQ.

In the next section, we propose a simple modification of the Deep Ensemble method that restores the guarantee of convergence to correct values of the QoI in the limit of a high number of samples.

\subsection{Importance Sampling}

To improve the approximation of expectations, one can use importance sampling. In this approach, rather than sampling directly from the posterior distribution of the weights \( p_{\text{pos}}(w \mid X, Y) \), we sample from an alternative distribution \( q(w) \). The expectation of a function \( f(y) \) given the input \( x \) is then adjusted by weighting the samples according to the ratio of the posterior to the sampling distribution(\cite{oh1992adaptive},\cite{neal2001annealed}, and \cite{tokdar2010importance}):
\[
\mathbb{E}[f(y) \mid x] = \int \mathbb{E}[f(y) \mid x, w] \frac{p_{\text{pos}}(w \mid X, Y)}{q(w)} q(w) \, dw.
\]
Here, \( w \) represents the neural network weights, \( X \) and \( Y \) denote the training data inputs and outputs respectively, and \( p_{\text{pos}}(w \mid X, Y) \) is the posterior distribution of the weights given the training data. \( q(w) \) is the alternative distribution from which we sample. In practice, for weights \( \{w_i\}_{i=1}^N \) sampled from \( q(w) \), the weighted empirical mean is used to approximate the expectation:
\[
\mathbb{E}[f(y) \mid x] \approx \frac{1}{N} \sum_{i=1}^N f(g_{w_i}(x)) \frac{p_{\text{pos}}(w_i \mid X, Y)}{q(w_i)}.
\]
In this expression, \( w_i \) are the sampled weights, \( g_{w_i}(x) \) is the output of the neural network with weights \( w_i \) given input \( x \), and \( N \) is the number of samples.

Next, consider the scenario where a deep ensemble is generated by iteratively applying gradient descent steps to an initially uniform distribution of weights. The weights evolve according to the rule
\[
w^{(k+1)} = \phi_k(w^{(k)}),
\]
where \( w^{(k)} \) represents the weights after \( k \) iterations, and \( \phi_k \) is the update function at step \( k \), typically representing a gradient descent update. The density of the weights \( q_k(w) \) after \( k \) steps is then updated by the push-forward operation:
\[
q_{k+1}(w) = \phi_k \sharp q_k(w) = q_k(\phi_k^{-1}(w)) \frac{1}{\det(\nabla \phi_k(\phi_k^{-1}(w)))}.
\]
Here, \( \phi_k^{-1}(w) \) is the inverse of the update function, mapping the weights \( w \) back to their previous state. The determinant \( \det(\nabla \phi_k(\phi_k^{-1}(w))) \) represents the change in volume in weight space caused by the update \( \phi_k \), where \( \nabla \phi_k \) is the Jacobian matrix of \( \phi_k \) with respect to \( w \).

To draw a connection to continuous dynamics, consider a small time step \( \Delta t \) where the update rule is approximated by
\[
\phi(w) = w + \Delta t \, \mu(w),
\]
with \( \mu(w) = -\nabla_w L(w) \) being the gradient of the loss function \( L(w) \) with respect to the weights. The Jacobian matrix \( \nabla \phi(w) \) is then given by
\[
\nabla \phi(w) = I + \Delta t \, \nabla_w \mu(w),
\]
where \( I \) is the identity matrix and \( \nabla_w \mu(w) \) is the Hessian matrix (second derivative) of \( \mu(w) \) with respect to the weights \( w \). The determinant of this matrix is
\[
\det(\nabla \phi(w)) = 1 + \Delta t \, \nabla_w \cdot \mu(w) + O((\Delta t)^2),
\]
where \( \nabla_w \cdot \mu(w) \) is the divergence of the gradient field \( \mu(w) \), representing the rate of change of the density under the flow induced by \( \mu(w) \). This leads to the push-forward approximation
\[
q_{k+1}(w) = q_k(w) \left(1 - \Delta t \, \nabla_w \cdot \mu(w)\right) + O((\Delta t)^2),
\]
where the density \( q_{k+1}(w) \) after the \( (k+1) \)-th update is adjusted by the divergence term \( \nabla_w \cdot \mu(w) \), reflecting how the probability density evolves under the gradient flow.

To relate this formula to the one obtained from the continuous time derivation in the appendix, let \( \phi(w) = w +  \Delta t \, \mu(w) \), where \( w \) represents the neural network weights, \( \Delta t \) is a small time step, and \( \mu(w) = -\nabla_w L(w) \) is the gradient of the loss function \( L(w) \) with respect to the weights. Then,
\[
\nabla \phi(w) = I + \Delta t \, \nabla \mu(w),
\]
where \( \nabla \phi(w) \) is the Jacobian matrix of \( \phi(w) \) with respect to \( w \), \( I \) is the identity matrix, and \( \nabla \mu(w) \) is the Hessian matrix, which is the second derivative of the function \( \mu(w) \) with respect to \( w \).

The determinant of the Jacobian matrix can be expressed as:
\begin{align*}
\mbox{det}(\nabla \phi(w)) &= 1 + \Delta t \, \mbox{tr}(\nabla \mu(w)) + O((\Delta t)^2) \\
&= 1 + \Delta t \, \nabla \cdot \mu(w) + O((\Delta t)^2),
\end{align*}
where \( \mbox{tr}(\nabla \mu(w)) \) is the trace of the Hessian matrix (i.e., the sum of its diagonal elements), and \( \nabla \cdot \mu(w) \) represents the divergence of the gradient field \( \mu(w) \), which is the sum of the partial derivatives of \( \mu(w) \) with respect to the components of \( w \).

Hence, the inverse of the determinant can be approximated by:
\[
\frac{1}{\mbox{det}(\nabla \phi(w))} = 1 - \Delta t \, \nabla \cdot \mu(w) + O((\Delta t)^2).
\]

On the other hand, consider the exponential term from the continuous time derivation:
\[
\exp  \left[ - \int_0^{\Delta t} (\nabla \cdot \mu)(w(t)) dt \right] = 1 - \Delta t \, \nabla \cdot \mu(w) + O((\Delta t)^2).
\]
Here, \( w(t) \) represents the weights as a function of time \( t \), and the integral \( \int_0^{\Delta t} (\nabla \cdot \mu)(w(t)) dt \) computes the cumulative effect of the divergence over the time interval \( [0, \Delta t] \). The exponential function approximates to the first-order term in \( \Delta t \), showing consistency with the discrete time result.

By combining ensemble methods with importance sampling, we can more accurately estimate uncertainty while keeping computational costs manageable. This approach involves calculating the importance of each ensemble member, effectively weighing their contributions to the overall uncertainty estimation. However, this process requires computing the trace of the Hessian matrix at each training epoch, which can be computationally demanding. To overcome this challenge, we use Hutchinson's trace estimator, a technique that approximates the Hessian trace efficiently, reducing the computational load. This refined method enables practical and precise uncertainty quantification in neural network models, even in complex applications like seismic inversion.

\section{Numerical Examples}
\label{experiments}

\subsection{Data Preparation}

In this research, we generated a total of 666 velocity model and seismic dataset pairs using Devito \cite{devito-api}. These datasets represent distinct geological realizations, created with the help of Gempy \cite{gmd-12-1-2019}. Out of the total datasets, 600 pairs were allocated for training, 30 pairs were set aside for validation, and the remaining 36 pairs were used for testing.

\subsection{Network Architecture and General Approach Implementation}

\begin{figure}[ht]
    \centering
    \includegraphics[width=0.8\textwidth]{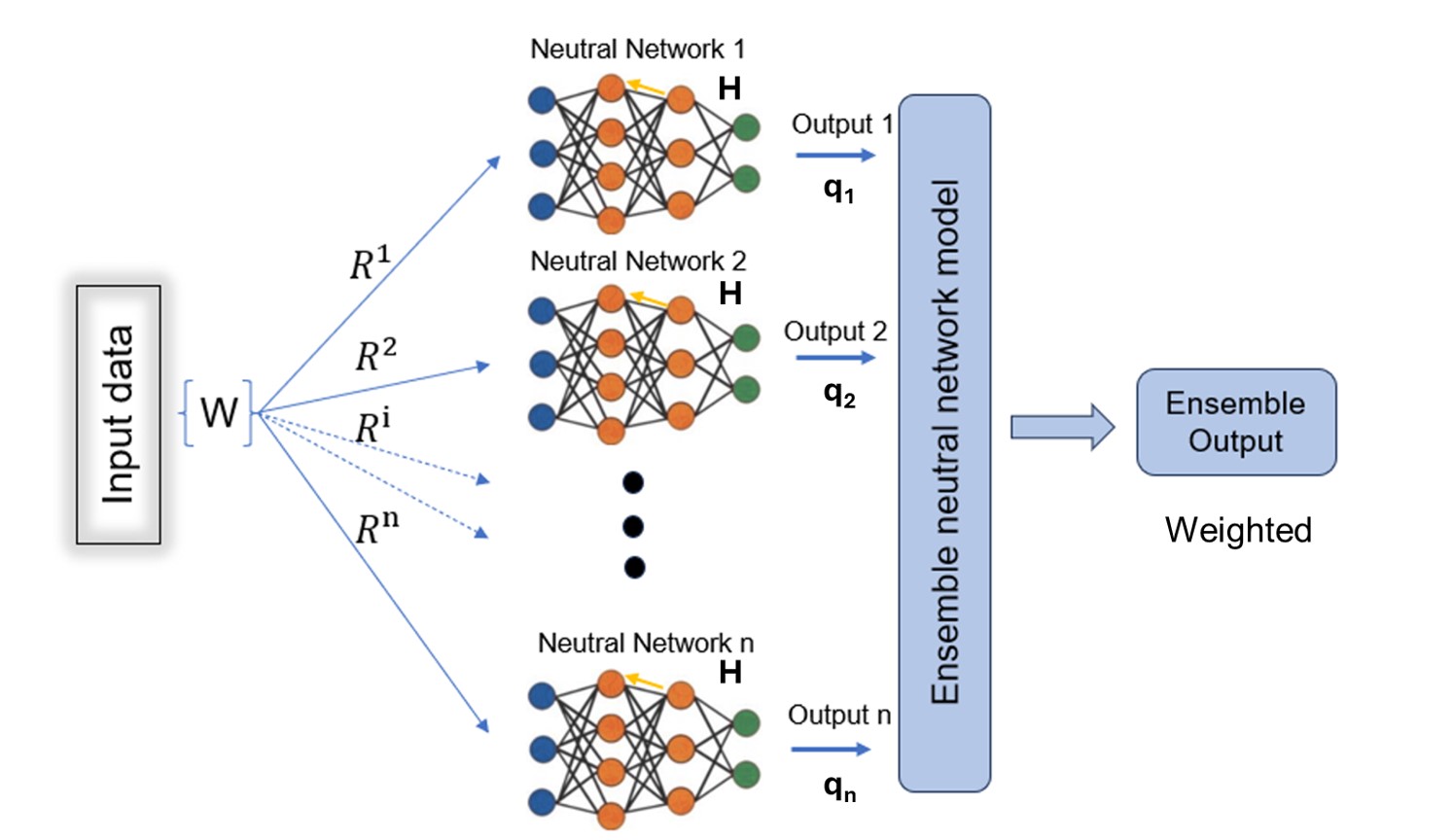}
    \caption{Diagram of the method approach.\label{architecture}}
\end{figure}

In this study, we deploy a modified version the U-Net architecture (originally introduce in \cite{unet2015} and use for seismic inversion in \cite{Gelboim23} among many), which is effective and reliable in regression tasks, to reconstruct velocity models from seismic data. Recognizing the inherent uncertainties associated with seismic data interpretation, we employ the Deep Ensemble technique to quantify these uncertainties without heavy computational cost. To enhance the efficiency and precision of our Bayesian inference process, we further integrate Importance Sampling. The general structure of the approach is depicted in Figure~\ref{architecture}. For this specific regression task, we employ the Mean Squared Error (MSE) loss and Mean Absolute Error (MAE) loss function.

\subsection{MC Droput Experiments}

To establish a baseline for our model's robustness and performance under uncertainty, we conducted a series of experiments utilizing Monte Carlo (MC) Dropout. The model was tested with varying dropout rates to assess their impact on predictive performance and uncertainty estimation.

MC Dropout was applied during both the training and prediction phases, allowing the model to sample from an approximate posterior distribution and providing a measure of predictive uncertainty. By systematically varying the dropout rate across multiple runs, we evaluated its influence on the model's accuracy and the resulting confidence intervals.

\begin{figure}[ht]
    \centering
    \includegraphics[width=\textwidth]{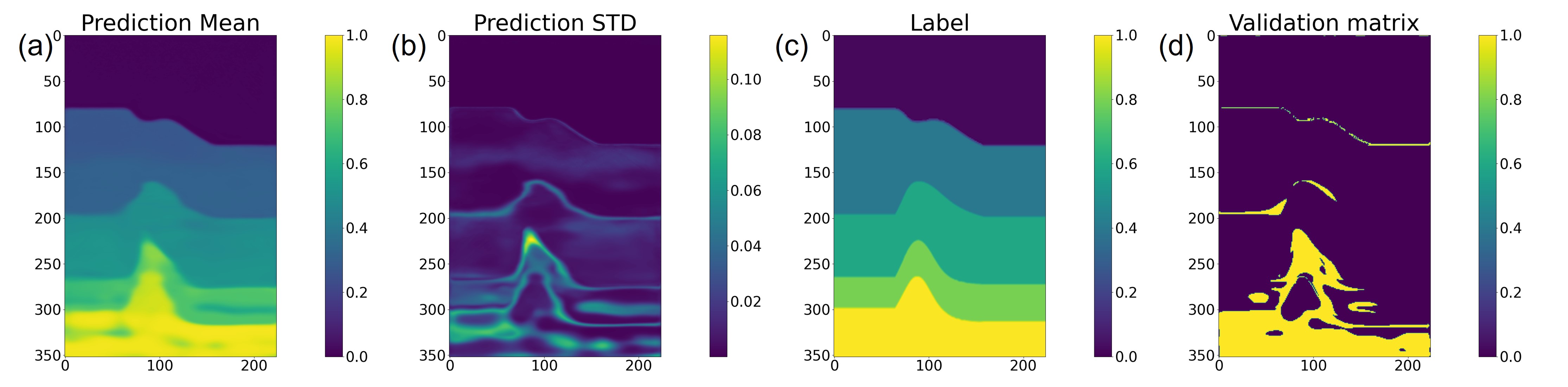}
    \caption{MCDropout results with dropout rate 0.2 and L2 misfit function.}
    \label{fig:mcdropout_0_2_l2_1200_1}
\end{figure}

Figure~\ref{fig:mcdropout_0_2_l2_1200_1} shows the results of applying the MC Dropout method to estimate uncertainty in the seismic inversion task, specifically using a dropout rate of 0.2, with an \( L_2 \) misfit function, and trained over 1200 epochs. The uncertainty is quantified based on 1500 prediction samples. Subfigure (a) shows the Prediction Mean, representing the average predicted velocity model across the 1500 prediction samples. This mean prediction provides an overall estimate of the subsurface structure, effectively smoothing out variability introduced by dropout during inference. Subfigure (b) illustrates the Prediction Standard Deviation (STD), which quantifies the uncertainty in the predictions. Higher standard deviations highlight regions where the model's predictions are less certain, possibly due to the complexity of the subsurface features or limitations in the data. This is particularly important in seismic inversion, where such areas might correspond to zones with significant subsurface variability or insufficient data constraints. Subfigure (c) displays the Label, which is the ground truth or reference model, allowing for a direct visual comparison between the predictions and the known subsurface structure. Finally, Subfigure (d) displays the Validation Matrix, a binary map that distinguishes regions based on the prediction uncertainty relative to the ground truth. Specifically, this matrix is critical for assessing the reliability of the MC Dropout method in determining where the model's predictions are dependable. In the Validation Matrix, blue regions indicate that the ground truth values fall within three times the standard deviation of the prediction mean, signifying areas where the model's uncertainty is within an acceptable threshold. Conversely, yellow regions highlight areas where the ground truth values lie outside this range, indicating zones of significant uncertainty. This analysis shows that while the Prediction Mean provides a reasonable approximation of the subsurface structure, the Prediction Standard Deviation (STD) uncovers areas with substantial uncertainty, particularly in more complex regions.

\begin{figure}[ht]
    \centering
    \includegraphics[width=\textwidth]{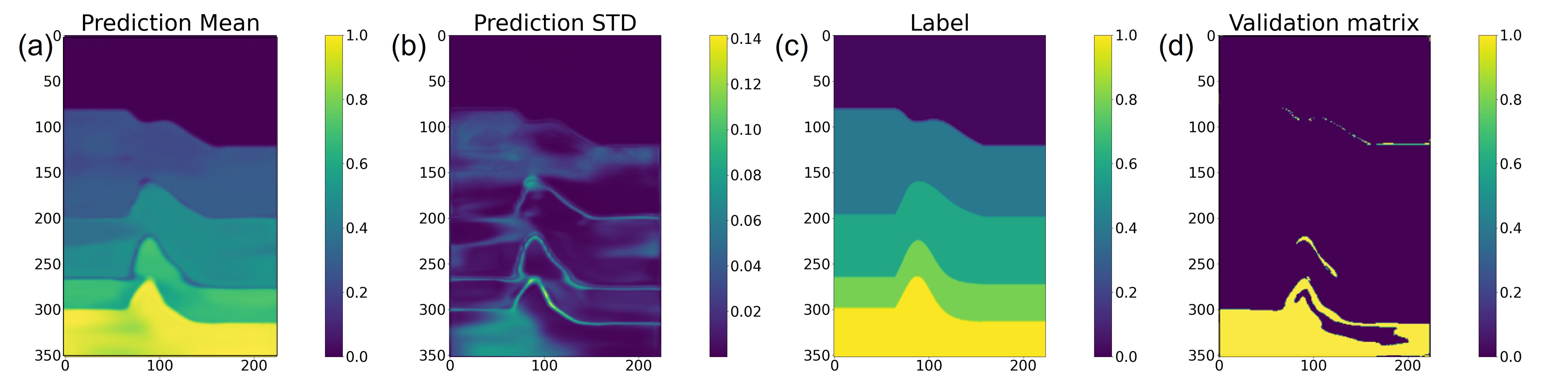}
    \caption{MCDropout results with dropout rate 0.2 and L1 misfit function.}
    \label{fig:mcdropout_0_2_l1_1200_2}
\end{figure}

Figure~\ref{fig:mcdropout_0_2_l1_1200_2} presents the results of applying the Monte Carlo Dropout (MC Dropout) method with a dropout rate of 0.2, using an \( L_1 \) misfit function, and trained over 1200 epochs. The overall approach remains consistent with the previous figure, but the key difference lies in the use of the \( L_1 \) norm for the misfit function. Subfigure (a) shows the Prediction Mean, providing an averaged estimate of the subsurface structure based on 1500 samples. Subfigure (b) displays the Prediction Standard Deviation (STD), which highlights the regions of uncertainty in the predictions. Compared to the \( L_2 \) misfit function, the results here demonstrate less uncertainty, particularly in areas with complex subsurface features, suggesting that the \( L_1 \) norm is more effective at capturing the true structure. Subfigure (c) represents the Label, which is the ground truth model, and Subfigure (d) shows the Validation Matrix, illustrating regions where prediction uncertainty is within acceptable limits. The improved clarity and reduced uncertainty in these results indicate that the \( L_1 \) misfit function provides better performance in this seismic inversion task compared to the \( L_2 \) misfit function.

\begin{figure}[ht]
    \centering
    \includegraphics[width=\textwidth]{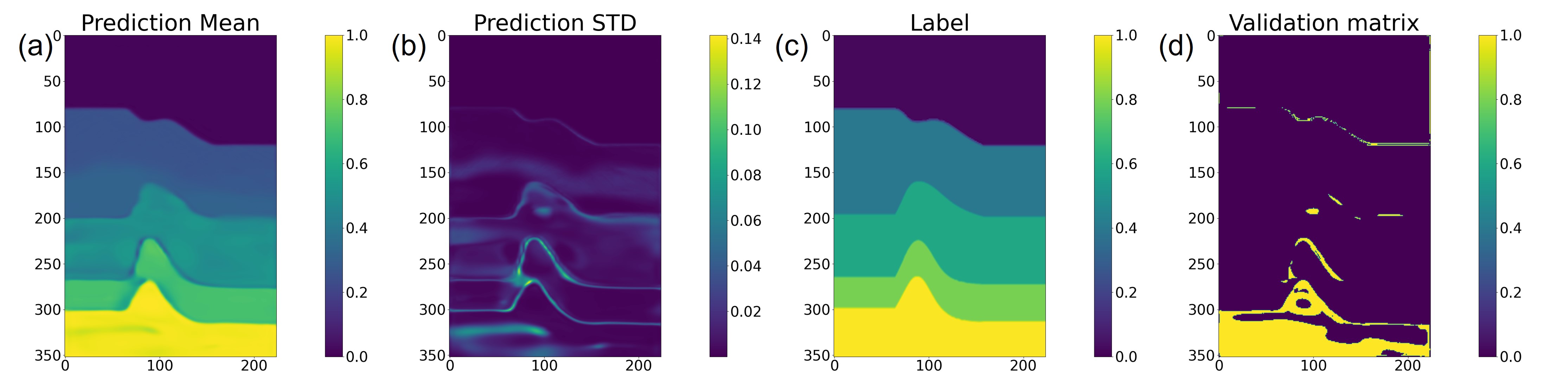}
    \caption{MCDropout results with dropout rate 0.3 and L1 misfit function.}
    \label{fig:dropout_0_3_l1_1200_3}
\end{figure}

\begin{figure}[ht]
    \centering
    \includegraphics[width=\textwidth]{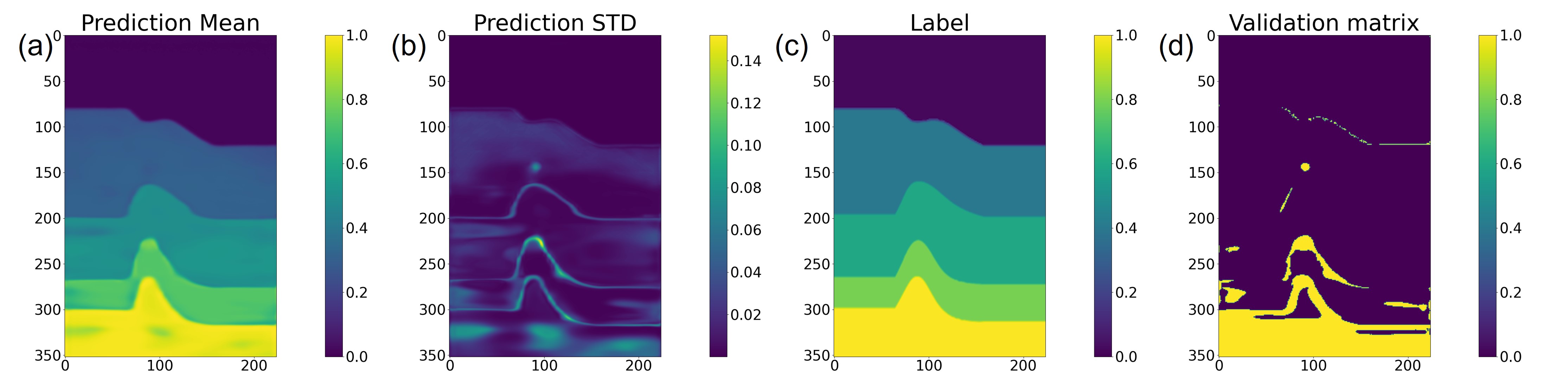}
    \caption{MCDropout results with dropout rate 0.4 and L1 misfit function.}
    \label{fig:dropout_0_40_l1_1200_4}
\end{figure}

\begin{figure}[ht]
    \centering
    \includegraphics[width=\textwidth]{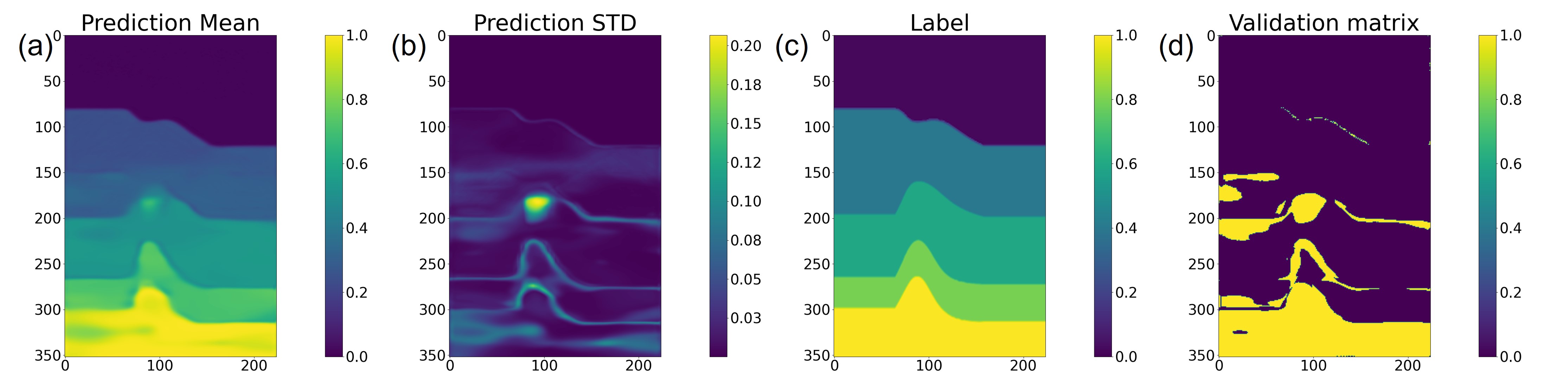}
    \caption{MCDropout results with dropout rate 0.5 and L1 misfit function.}
    \label{fig:mcdropout_0_5}
\end{figure}

Figure~\ref{fig:dropout_0_3_l1_1200_3}, Figure~\ref{fig:dropout_0_40_l1_1200_4} and Figure~\ref{fig:mcdropout_0_5} are the MCDropout results with dropout rate to be 0.3, 0.4, and 0.5, respectively. The three figures demonstrate the application of Monte Carlo Dropout (MC Dropout) using an \( L_1 \) misfit function and trained over 1200 epochs. As the dropout rate increases, several important trends emerge that highlight the characteristics of MC Dropout in uncertainty quantification. The Prediction Mean across all three dropout rates provides a consistent and smooth estimate of the subsurface structure, though with slightly more smoothing at higher dropout rates, which may result in less detailed predictions in regions with complex subsurface features. The Prediction Standard Deviation (STD), representing the uncertainty in the predictions, increases noticeably with higher dropout rates. This reflects a core feature of MC Dropout in UQ—the introduction of more variability in the model with higher dropout rates leads to higher predicted uncertainty, particularly in areas with significant structural changes in the subsurface. The Validation Matrix further emphasizes this trend, showing larger areas of higher uncertainty as the dropout rate increases. This underscores the trade-off inherent in adjusting the dropout rate: while lower rates (e.g., 0.3) yield more confident predictions with less variability, higher rates (e.g., 0.5) capture greater uncertainty, which may better represent the underlying data's variability but can also result in less precise predictions.

\subsection{Deep Ensemble Experiments}

Next, we conducted a series of Deep Ensemble experiments. For each ensemble member, we initialized the weights differently using Kaiming initialization. To further improve robustness, we set the dropout rate to 0.2 during training, adding an additional layer of uncertainty estimation. The prior distribution of the weights was assumed to be uniform, ensuring that the ensemble members explore a diverse set of model configurations. This approach not only helps in capturing a broader range of potential solutions but also contributes to a more reliable uncertainty quantification in the resulting predictions.

\begin{figure}[ht]
    \centering
    \includegraphics[width=\textwidth]{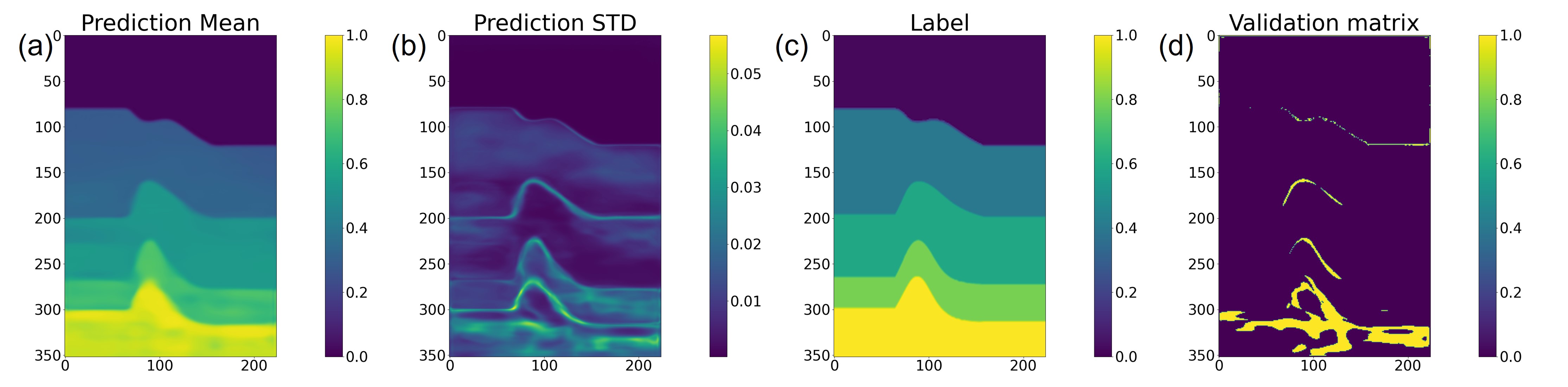}
    \caption{Deep Ensemble results with 10 Samples, Dropout Rate 0.2, and 800 Epochs.}
    \label{Deep Ensemble results with 10 Samples, Dropout Rate 0.2, and 800 Epochs}
\end{figure}

Figure \ref{Deep Ensemble results with 10 Samples, Dropout Rate 0.2, and 800 Epochs} illustrates the results of applying a deep ensemble method with 10 ensemble samples, each trained with a dropout rate of 0.2 over 800 epochs. Subfigures have the same meaning as in the previous section. The results suggest that the ensemble approach, combined with a moderate dropout rate, effectively captures the uncertainty in complex subsurface areas while maintaining accurate overall predictions. In particular, when comparing with MC Dropout results, the validation matrix in Figure \ref{Deep Ensemble results with 10 Samples, Dropout Rate 0.2, and 800 Epochs} contains far less unacceptable elements.

\begin{figure}[!ht]
    \centering
    \includegraphics[width=\textwidth]{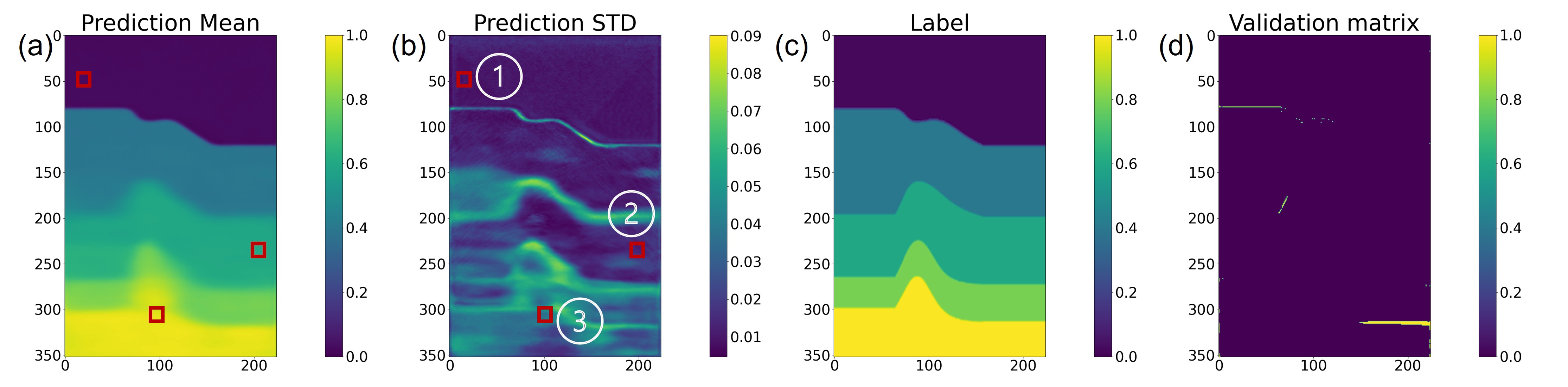}
    \caption{Deep Ensemble results with 20 Samples, Dropout Rate 0.2, and 200 Epochs.}
    \label{fig:ensemble_20_200_8}
\end{figure}

\begin{figure}[!ht]
    \centering
    \includegraphics[width=\textwidth]{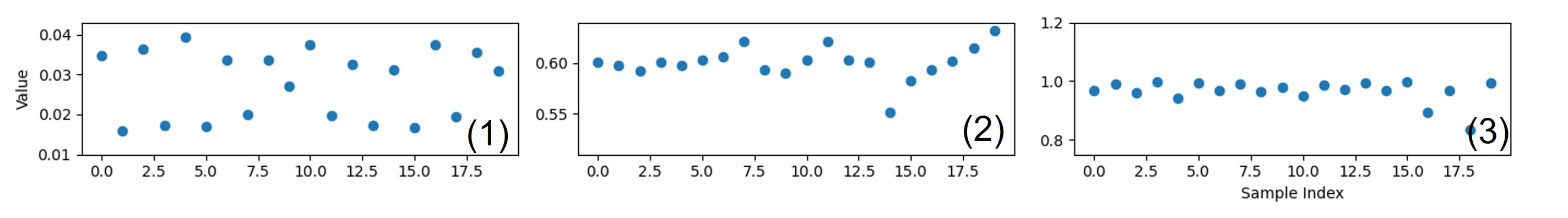}
    \caption{Distribution of Posterior 20 Samples at Selected Locations.}
    \label{fig:ensemble_20_l1_200_sub}
\end{figure}

\begin{figure}[!ht]
    \centering
    \includegraphics[width=0.6\textwidth]{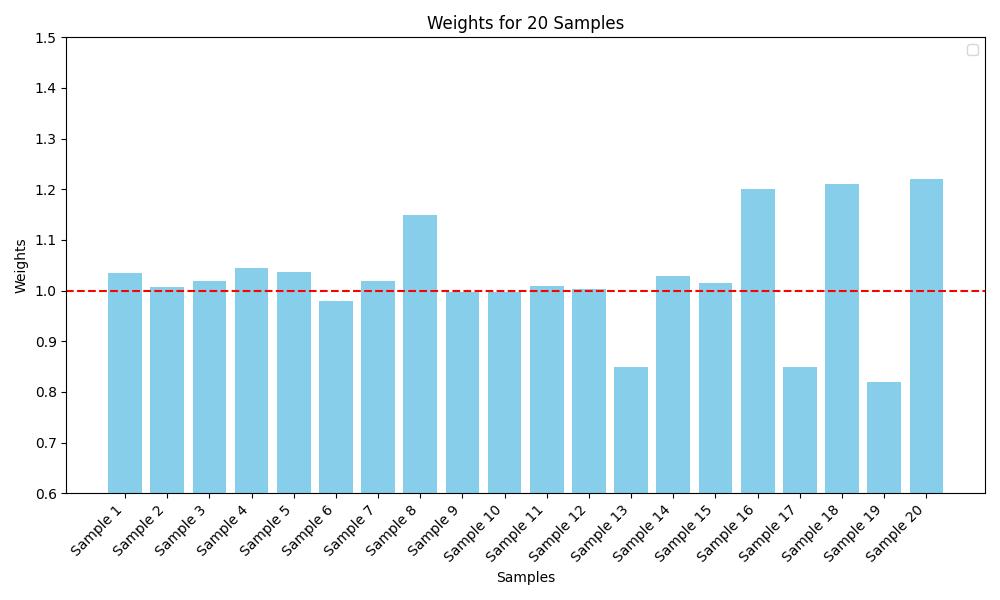}
    \caption{Weights for the 20 samples.}
    \label{fig:sample_weights}
\end{figure}

Considering the computational burden, we did not use 800 epochs for all the training; instead, the later training was mostly done with 200 or 300 epochs.
The presented figures (Figure~\ref{fig:ensemble_20_200_8} and \ref{fig:ensemble_20_l1_200_sub}) illustrate the results of a deep ensemble method using 20 ensemble samples, trained over 200 epochs.
Figure \ref{fig:ensemble_20_l1_200_sub} provides scatter plots of the posterior distributions at three specific locations marked by red squares in the first figure. These scatter plots give a detailed view of the variability and uncertainty captured by the ensemble at each location. Each plot reflects the distribution of posterior samples, offering insights into how the model's uncertainty manifests at these critical points. The dispersion of points in the scatter plots indicates the range of possible values that the model considers plausible, with broader distributions suggesting higher uncertainty. The weights for each samples is shown in Figure~\ref{fig:sample_weights}.

\begin{figure}[!ht]
    \centering
    \includegraphics[width=\textwidth]{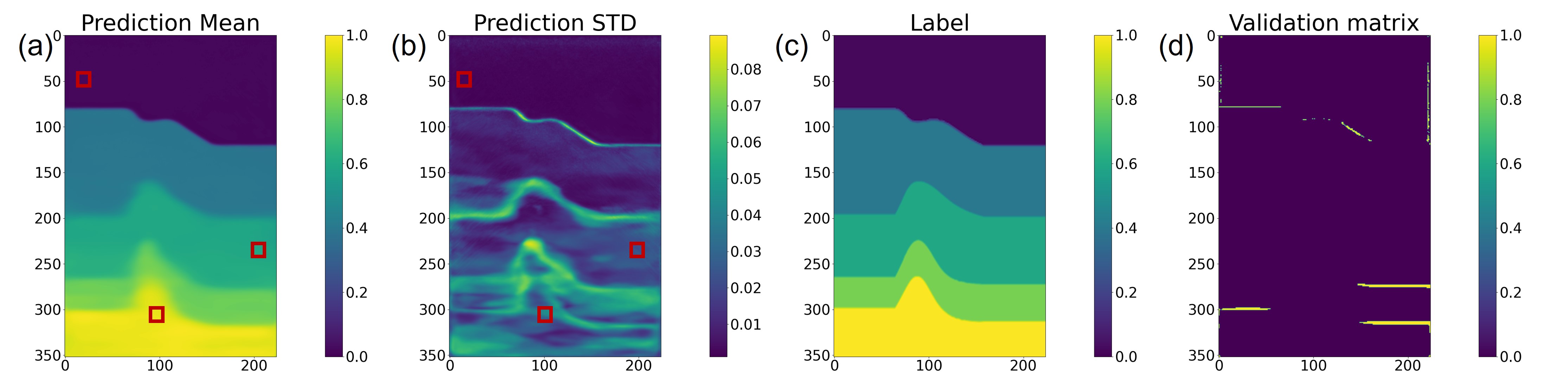}
    \caption{Deep Ensemble results with 20 Samples, Dropout Rate 0.2, and 300 Epochs.}
    \label{fig:ensemble_20_l1_300_9}
\end{figure}

\begin{figure}[!ht]
    \centering
    \includegraphics[width=\textwidth]{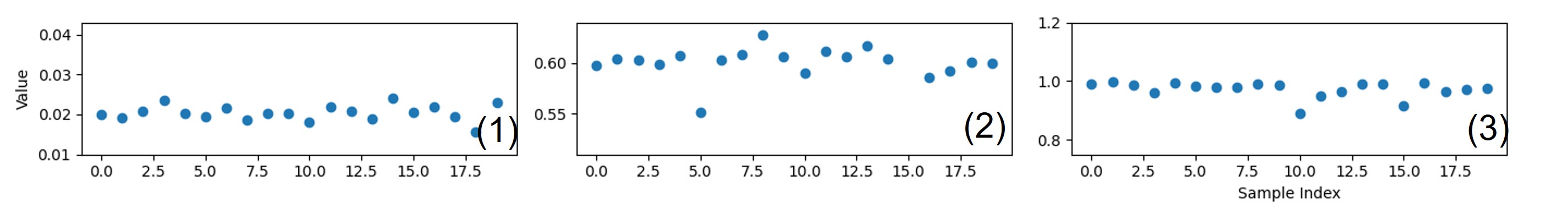}
    \caption{Distribution of Posterior 20 Samples at Selected Locations.}
    \label{fig:ensemble_20_l1_300_9_1}
\end{figure}

The figures (Figure~\ref{fig:ensemble_20_l1_300_9} and \ref{fig:ensemble_20_l1_300_9_1}) demonstrate the effects of increasing the number of training epochs from 200 to 300 while maintaining an ensemble size of 20. Compared to the previous figure, which utilized 200 epochs, the additional training iterations have contributed to a more refined and confident model output. This improvement is particularly noticeable in the Prediction Standard Deviation (STD), where the uncertainty patterns have become more localized and less pronounced. The reduction in uncertainty suggests that the model has converged more effectively, particularly in regions where the subsurface structure is complex.

Furthermore, the Validation Matrix in the 300-epoch figure indicates a closer alignment with the ground truth, especially in areas where the previous figure showed higher uncertainty. This suggests that the extended training has allowed the model to better capture the underlying patterns, reducing the number of regions with significant prediction errors. The scatter plots of posterior distributions at selected locations (as indicated by red squares) show tighter distributions, reflecting a reduction in variability and a stronger consensus among the ensemble members. This tighter clustering indicates that with more epochs, the ensemble models are converging towards similar predictions, enhancing the reliability of the overall uncertainty quantification.

\begin{figure}[ht]
    \centering
    \includegraphics[width=\textwidth]{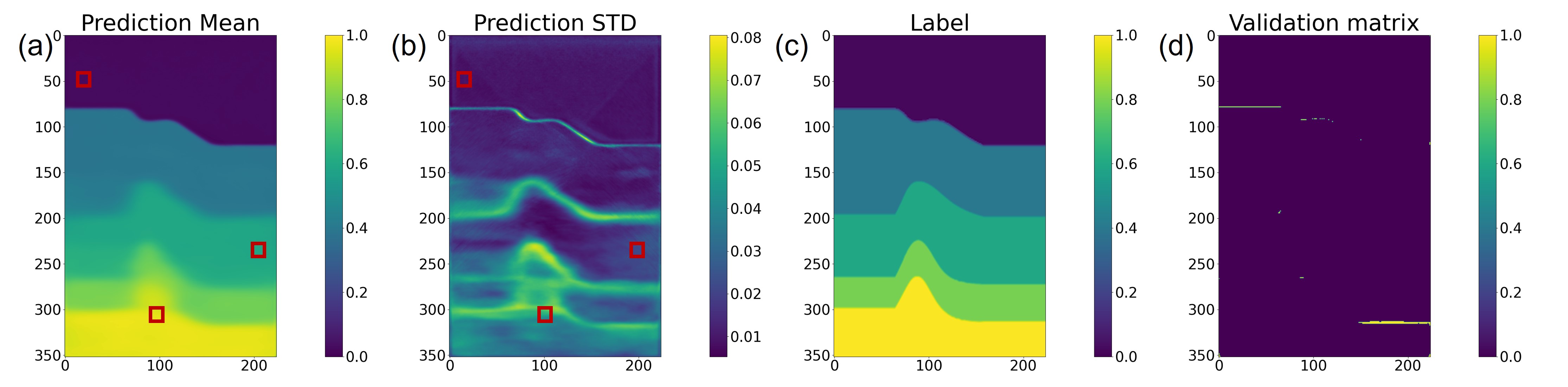}
    \caption{Deep Ensemble results with 50 Samples, Dropout Rate 0.2, and 300 Epochs.}
    \label{fig:figure_10}
\end{figure}

\begin{figure}[ht]
    \centering
    \includegraphics[width=\textwidth]{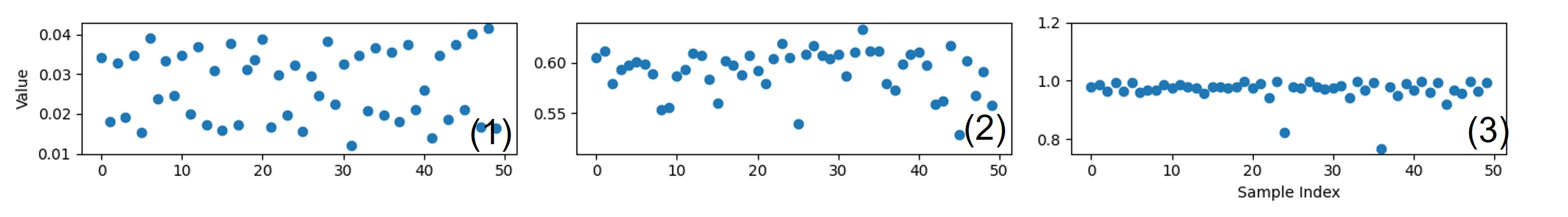}
    \caption{Distribution of Posterior 50 Samples at Selected Locations.}
    \label{fig:10_1}
\end{figure}


The figures (\ref{fig:figure_10} and \ref{fig:10_1}) present the results of a deep ensemble method applied with 50 ensemble samples, trained over 300 epochs. Compared to previous figures that utilized fewer ensemble samples or fewer epochs, the increased ensemble size and extended training period significantly enhance the model's ability to capture and quantify uncertainty.

In the main figure, the Prediction Mean (subfigure a) continues to display a smooth and consistent estimate of the subsurface structure, similar to the previous figures. However, the additional ensemble members and extended training lead to even more refined and confident predictions, particularly in regions with complex subsurface features. The Prediction Standard Deviation (STD) (subfigure b) exhibits further localized and reduced uncertainty patterns compared to the figures with fewer ensemble samples or epochs. The decreased STD values indicate that the model is achieving higher consensus among the ensemble members, leading to a more precise understanding of the underlying geological structures.  The scatter plots show a tighter clustering of points compared to previous figures, indicating a higher level of agreement among the ensemble members. This increased agreement reflects the model's ability to converge towards a more consistent and reliable estimate of uncertainty with the larger ensemble size.

\section{Conclusions}

In this study, we explored the application of advanced uncertainty quantification techniques in seismic inversion, focusing on both Monte Carlo Dropout and Deep Ensemble methods. Our primary objective was to enhance the accuracy and reliability of UQ in neural network-based models by leveraging these techniques.

The MC Dropout approach was shown to effectively quantify uncertainty by introducing stochastic variability during both training and inference. This method provided a flexible way to estimate model uncertainty, particularly in regions with complex subsurface structures. However, while MC Dropout is computationally efficient and easy to implement, it does have limitations in terms of the precision of the uncertainty estimates, especially in models with a large number of parameters.
To address these limitations, we implemented a Deep Ensemble method, which improves UQ by training multiple models with different initializations. This method captured a broader range of possible outcomes, leading to more accurate and reliable predictions. By increasing the ensemble size and training duration, we demonstrated that the ensemble method could significantly reduce prediction uncertainty, particularly in challenging subsurface areas.

Furthermore, by integrating importance sampling with the ensemble method, we were able to further refine our UQ estimates. Importance sampling allowed us to weigh the contributions of each ensemble member according to their likelihood, ensuring that the most plausible models had a greater influence on the final predictions. This approach provided a more accurate and computationally efficient way to estimate the posterior distribution of the model parameters.

Overall, our findings indicate that the combination of Deep Ensemble methods with importance sampling represents a powerful strategy for enhancing UQ in seismic inversion tasks. By improving the accuracy and reliability of uncertainty estimates, this approach enables more informed decision-making in geophysical exploration and resource management. The results of this study underscore the importance of utilizing advanced UQ techniques to better capture the complexities and uncertainties inherent in seismic inversion, ultimately leading to more robust and dependable models.

\section{Discussions}

This study has addressed the challenge of uncertainty quantification in seismic inversion, a problem characterized by its high dimensionality, non-linearity, multimodality, and mixed-determined nature. These intrinsic complexities make seismic inversion a particularly difficult task, where accurately capturing the range of possible subsurface models is crucial for reliable decision-making. Our approach, combining Monte Carlo Dropout and Deep Ensemble methods, aimed to tackle these challenges by improving the robustness and accuracy of UQ.

Seismic inversion is inherently a high-dimensional problem, where the model parameters are numerous and interact in complex ways. This high dimensionality increases the computational burden significantly, especially when multiple models or ensemble methods are employed to quantify uncertainty. Moreover, the inverse problem is non-linear, meaning small changes in input data can lead to disproportionately large changes in the model parameters. This non-linearity complicates the inversion process and requires methods that can effectively navigate the rugged, multi-peaked (multimodal) posterior landscape where multiple solutions may fit the observed data equally well.

In addressing these challenges, ensemble methods have proven particularly valuable. By generating multiple models through different initial conditions and parameter settings, deep ensemble methods are capable of exploring the multimodal nature of the solution space more thoroughly than single-model approaches. Each ensemble member represents a plausible solution, and collectively, they provide a comprehensive view of the uncertainty associated with the inversion process.

However, the increased computational demand associated with running large ensembles cannot be ignored. High-dimensional, non-linear, and multimodal problems, such as seismic inversion, require extensive computational resources to generate and evaluate multiple models. The mixed-determined nature of the problem—where some parameters are well-constrained by the data while others are not—further complicates this task, as it demands a careful balancing of computational efforts to ensure that the ensemble adequately captures the range of uncertainty without becoming prohibitively expensive. Additionally, it is important to note that the computational cost of calculating the \text{importance} through methods like Hessian trace estimation during practical implementation is also substantial. This added computational burden means that before fully committing to this approach, it is crucial to test whether this method is truly suitable and efficient for the specific problem at hand. If the computational demands outweigh the benefits, alternative approaches or optimizations may need to be considered to ensure the method's feasibility in practical applications.

To mitigate these computational challenges, we integrated importance sampling into the ensemble framework. Importance sampling allows us to prioritize ensemble members that are more likely to represent the true posterior distribution, thereby reducing the number of required samples without sacrificing the accuracy of UQ. This method proved to be effective in refining our uncertainty estimates, especially in the context of a high-dimensional and complex problem like seismic inversion. It enabled us to focus computational resources on the most informative parts of the solution space, thereby enhancing the efficiency and precision of the ensemble approach.

In summary, while the application of Deep Ensemble methods in high-dimensional, non-linear, multimodal, and mixed-determined inverse problems like seismic inversion is computationally demanding, the benefits in terms of improved UQ are substantial.

\section{Acknowledgments}

The authors thank TotalEnergies EP Research \& Technology USA, for supporting this work and allowing its publication.\\

\noindent{\bf \large Appendix}
\appendix
\section{Continuous-Time Formulation of Deep Ensembles}

The Deep Ensemble can also be considered as generated from a gradient flow on an initially uniform distribution of weights. The weight evolution is described by the differential equation
\[
\frac{dw}{dt} = \mu(w),
\]
where \( \mu(w) = -\nabla_w L(w) \) is the negative gradient of the loss function. The resulting probability density function \( q_t(w) \), indexed by time \( t \), satisfies the Fokker-Planck equation(\cite{jordan1998variational}):
\[
\frac{\partial q_t}{\partial t} + \nabla_w \cdot (\mu(w) q_t(w)) = 0,
\]
with the initial condition \( q_0(w) = \text{const} \).

This partial differential equation can be solved using the method of characteristics. By defining the material derivative \( \frac{D}{Dt} = \frac{\partial}{\partial t} + \mu(w) \cdot \nabla_w \), the density function along a characteristic curve satisfies the ordinary differential equation:
\[
\frac{D q_t}{D t} + (\nabla_w \cdot \mu(w)) q_t = 0.
\]
Therefore, the density \( q_t(w(t)) \) evolves as
\[
q_t(w(t)) = q_0(w(0)) \exp\left[ -\int_0^t (\nabla_w \cdot \mu(w(t'))) dt' \right].
\]

\bibliographystyle{plain}  
\bibliography{references}  



\end{document}